\documentclass[12pt,longbibliography]{article}
\usepackage{amsmath,amssymb,amsthm,amsxtra,overpic,bbm,bm,epsfig,subfigure}
\usepackage{multirow} 
\usepackage{booktabs}
\usepackage{hyperref}
\usepackage{mathrsfs}
\usepackage{graphicx}
\usepackage{multirow}
\usepackage{makecell}
\usepackage{verbatim}
\usepackage{color}
\usepackage{comment}
\usepackage{epstopdf}
\usepackage{float}
\usepackage{amsmath}
\usepackage{amssymb}
\usepackage{tikz}
\usepackage{wrapfig}
\usepackage{tikz-feynman}
\usepackage{verbatim}
\numberwithin{equation}{section}
\usepackage{cite}
\usepackage{hyperref}
\usepackage{url}
\usepackage{slashed,stmaryrd}

\newcommand*{\dif}{\mathop{}\!\mathrm{d}}

\begin{document}
	\vspace{0.2cm}
\begin{center}
	{\Large\bf Probing neutrino magnetic moments and the Xenon1T excess with coherent elastic solar neutrino scattering}
\end{center}
\vspace{0.2cm}

\begin{center}
	{\bf Yu-Feng Li}~$^{a,~b}$~\footnote{E-mail: liyufeng@ihep.ac.cn},
	{\bf Shuo-yu Xia}~$^{a,~b}$~\footnote{E-mail: xiashuoyu@ihep.ac.cn (corresponding author)}
	\quad
	\\
	\vspace{0.2cm}
	{\small
		$^a$Institute of High Energy Physics, Chinese Academy of Sciences, Beijing 100049, China\\
		$^b$School of Physical Sciences, University of Chinese Academy of Sciences, Beijing 100049, China}
\end{center}

\begin{abstract}
The detection potential of the active neutrino magnetic moment ($\nu_{a}$MM) and the active sterile transition magnetic moment ($\nu_{s}$MM) in the solar neutrino CE$\nu$NS process from future dark matter direct detection experiments is studied and compared
with the respective allowed range to explain the Xenon1T excess.
We find that the sensitivity to the $\nu_{a}$MM approaches to the level between $10^{-10}\mu_B$ and $10^{-11}\mu_B$, which is dominantly limited by the detection threshold.
On the other hand, the future solar neutrino CE$\nu$NS detection would be powerful in probing the $\nu_{s}$MM for the sterile neutrino mass below 10 MeV, which can unambiguously test the $\nu_{s}$MM explanation of the Xenon1T excess.
The sensitivity in the general framework with both $\nu_{a}$MM and $\nu_{s}$MM contributions is also derived.
\end{abstract}

\newpage
\section{Introduction}
The observation of coherent elastic neutrino-nucleus scattering~\cite{Freedman:1973yd ,Freedman:1977xn} (CE$\nu$NS) by the COHERENT Collaboration at the spallation neutron source at Oak Ridge National Laboratory~\cite{COHERENT:2017ipa,COHERENT:2020iec,Akimov:2021dab} has unlocked a new and powerful tool to explore various physics phenomena in very diverse research fields including particle physics~\cite{Coloma:2017ncl,Liao:2017uzy,Papoulias:2017qdn,Denton:2018xmq,AristizabalSierra:2018eqm}, nuclear physics~\cite{Cadeddu:2017etk,Papoulias:2019lfi}, astrophysics~\cite{Giunti:2014ixa,Giunti:2015gga,Abdullah:2022zue} and cosmology~\cite{Brdar:2020quo}. The CE$\nu$NS process provides an innovative access to investigate the nuclear neutron density distributions~\cite{Cadeddu:2017etk,Cadeddu:2020lky,Cadeddu:2021ijh,Ciuffoli:2018qem,Papoulias:2019lfi,Coloma:2020nhf} as well as the weak mixing angle~\cite{Papoulias:2017qdn,Cadeddu:2018izq} and the neutrino electromagnetic properties~\cite{Cadeddu:2018dux,Cadeddu:2019eta,Kim:2021lun}. Besides, CE$\nu$NS can be a new play ground for the new physics beyond the Standard Model (SM)~\cite{Coloma:2017ncl,Liao:2017uzy,Papoulias:2017qdn,Denton:2018xmq,AristizabalSierra:2018eqm,Giunti:2014ixa,Giunti:2015gga,Brdar:2020quo,Cadeddu:2018dux,Cadeddu:2019eta,Kim:2021lun,Giunti:2019xpr,Corona:2022wlb,Li:2022jfl}.

CE$\nu$NS experiments based on the neutrino fluxes from man-made sources including the spallation neutron source~\cite{Barbeau:2021exu,Baxter:2019mcx} and nuclear reactors~\cite{CONUS:2020skt,CONUS:2021dwh,CONNIE:2019xid,CONNIE:2021ngo,Coloma:2022avw,Sierra:2022ryd} are promising on providing competitive constraints on a series of the new physics theories.
On the other hand,
there have been intensive interests in the observation of astrophysical neutrino fluxes through the CE$\nu$NS process, which could be a new probe to the supernovae~\cite{Lang:2016zhv,Pattavina:2020cqc,RES-NOVA:2021gqp,Raj:2019sci,Raj:2019wpy,Huang:2021enl}, the collapsing supermassive stars~\cite{Munoz:2021sad} and the primordial black holes~\cite{Calabrese:2021zfq}. Solar neutrinos, which have already been observed with the charged-current (CC)~\cite{GALLEX:1992gcp,GNO:2000avz,Abazov:1991rx,SAGE:1999uje,Cleveland:1998nv,SNO:2001kpb}, neutral current (NC)~\cite{SNO:2002tuh,SNO:2003bmh,SNO:2008gqy} and elastic scattering (ES)~\cite{Kamiokande-II:1989hkh,Super-Kamiokande:2001ljr,Borexino:2007kvk,Borexino:2008fkj,Collaboration:2011nga,BOREXINO:2014pcl,BOREXINO:2020aww} channels, and provide a stable flux for CE$\nu$NS experiments as one of the most intensive natural neutrino sources at the Earth. Meanwhile, the recoil energy region of the solar neutrino CE$\nu$NS process are located at the same target range of direct detection (DD) experiments
of weakly interacting massive particles (WIMPs) as the promising dark matter candidate, making it an important component of the neutrino floor~\cite{Boehm:2018sux,Gonzalez-Garcia:2018dep,Papoulias:2018uzy,Chao:2019pyh,AristizabalSierra:2021kht,OHare:2020lva}, as well as a suitable target signal for detection.

Recently, the Xenon1T Collaboration released the recoiled electron spectrum in the dark matter searches and observed an unexplained excess of 53 $ \pm $ 15 events above the expected background at a statistical significance of around 3$\sigma$~\cite{XENON:2020rca}. 
This excess is reported to be consistent with the solar neutrino ES process due to the active neutrino magnetic moment ($\nu_{a}$MM) in the region of (1.4, 2.9) $\times 10^{-11} \mu_B$, which is close to the current best limit reported by the Borexino Collaboration~\cite{PhysRevD.96.091103}.
However it is stringently limited by indirect constraints from analyses of white dwarfs~\cite{C_rsico_2014}, red-giants~\cite{Capozzi:2020cbu} and globular clusters~\cite{2019arXiv191010568A}, which have constrained the $\nu_{a}$MM at the level of $10^{-12} \mu_B$. 
Another appealing solution is the active-sterile transition magnetic moments ($\nu_{s}$MM), also known as neutrino dipole portal~~\cite{Magill:2018jla,Coloma:2017ppo,Schwetz:2020xra,Masip:2012ke,Bertuzzo:2018itn}, 
which is capable to explain the Xenon1T excess~\cite{Shoemaker:2018vii,Shoemaker:2020kji,Plestid:2020vqf,Papoulias:2019xaw,Balantekin:2013sda}, and still compatible with the laboratory and astrophysical bounds~\cite{Brdar:2020quo}. At present, dark matter DD experiments are entering the phase of the multi-ton scale and experiments like PandaX-4T~\cite{PandaX:2018wtu}, DARWIN~\cite{DARWIN:2016hyl} and DarkSide-20k~\cite{DarkSide-20k:2017zyg} can be an effective platform to detect CE$\nu$NS process due to lower energy threshold and higher target material mass. In addition, experiments employing germanium or silicon as the target material like SuperCDMS~\cite{SuperCDMS:2016wui} and EDELWEISS~\cite{EDELWEISS:2017uga} have achieved an extremely low threshold, which will significantly improve the efficiency to probe possible new physics beyond the SM. Therefore, CE$\nu$NS process in DD experiments can be another access to explore the explanations for the Xenon1T excess.

In this work, we are going to study the detection potential of the $\nu_{a}$MM and $\nu_{s}$MM  with the solar neutrino CE$\nu$NS detection in the next generation and future DD experiments. we present the sensitivity of the $\nu_{a}$MM and $\nu_{s}$MM, and compare
with the allowed range to explain the Xenon1T excess and other experimental constraints.
We find that the sensitivity to the $\nu_{a}$MM lies in the level of [$10^{-10}$, $10^{-11}$]$\mu_B$, which is dominantly limited by the detection threshold. Only the future silicon detector could match the allowed region to explain the Xenon1T excess. On the other hand, the solar neutrino CE$\nu$NS would be powerful in probing the $\nu_{s}$MM for the sterile neutrino mass below 10 MeV. All the considered scenarios can unambiguously test the $\nu_{s}$MM explanation of the Xenon1T excess with at least one order of magnitude better detection sensitivities.
We also derive the detection potential in the general framework with both $\nu_{a}$MM and $\nu_{s}$MM contributions in the CE$\nu$NS process.

The plan of this work is as follows. In Section II, we briefly describe the theoretical calculation of the CE$\nu$NS cross section in the presence of the $\nu_{a}$MM and $\nu_{s}$MM. In Section III, we illustrate the setup of the experimental scenarios and statistical analysis method. In Section IV, we present the numerical constraints. Finally we give the concluding remarks in Section V.

\section{Theoretical Framework}

To begin with, we illustrate the theoretical calculation of the CE$\nu$NS cross section under the SM and in the presence of the neutrino magnetic moment ($\nu$MM). The cross section in this case can be written as
\begin{equation}
\frac{\dif{\sigma}}{\dif{T}}(E_{\nu},T)=\frac{\dif{\sigma_{\mathrm{SM}}}}{\dif{T}}(E_{\nu},T)+\frac{\dif{\sigma_{\mathrm{\nu MM}}}}{\dif{T}}(E_{\nu},T)\,,
\end{equation}
where $ \dif{\sigma_{\mathrm{SM}}}/\dif{T} $ is the differential cross section of the CE$\nu$NS process between a neutrino with the energy $E_{\nu}$ and a nucleus with $Z$ protons and $N$ neutrons in the SM and can be written as
\begin{equation}
	\begin{aligned}
		\frac{\dif{\sigma_{\mathrm{SM}}}}{\dif{T}}(E_{\nu},T)=&\frac{G_{F}^2  M}{\pi}\left(1-\frac{MT}{2E_{\nu}^2}\right)[g_{V}^{p} Z F_{Z}\left(q^2\right) +g_{V}^{n} N F_{N}\left( q^2\right) ]^{2}\,,
	\end{aligned}
	\label{eq:csSM} 
\end{equation}
where $T$ is the kinetic energy of nuclear recoil, $M$ is the nucleus mass and $G_{F}$ is the Fermi constant. $g_{V}^{p}$ and $g_{V}^{n}$ are the vector neutrino-proton and neutrino-neutron couplings respectively, which can be written as
\begin{equation}
	g_{V}^{p}=-2\sin^{2}\theta_{W}+\frac{1}{2}\simeq0.0229,\quad\quad g_{V}^{n}=-\frac{1}{2}\,,
\end{equation}
where $\theta_{W}$ is the weak mixing angle at low momentum transfer and the radiative corrections have been neglected~\cite{Cadeddu:2020lky}. Meanwhile, $ F_{Z}\left(q^2\right) $ and $ F_{N}\left(q^2\right) $ are the form factors of nucleon distributions in the nucleus for proton and neutron respectively and $q^2=|\vec{q}|^{2}=2MT$ is the square of momentum transfer. In this work, we use the Helm parameterization~\cite{Helm:1956zz} for the form factors of all considered nuclei and neglect the difference between proton and neutron form factors. The Helm proton form factor can be written as 
\begin{equation}
	F_{Z}\left({q}^{2}\right)=3 \frac{j_1\left(q R_{Z}\right)}{q R_{Z}}e^{-{q^{2} s^{2}}/{2}}\,,
\end{equation}
where the surface thickness $s=0.9\;{\rm fm}$ and the proton radii $ R_{Z} $ from Ref.~\cite{PhysRevLett.125.141301} have been employed for the target nuclei in the following calculation. Note that the cross section will be weighted averaged according to the natural abundance if there are several stable isotopes for the target material.


The general form of the CE$\nu$NS cross section induced by the $\nu$MM for a neutrino mass eigenstate with the mass $m_i$ scattered to the mass eigenstate with mass $ m_j $ can be written in a similar form as that of the electron scattering process~\cite{Balantekin:2013sda}:
\begin{equation}
\frac{\dif \sigma^{ij}_{\nu {\rm MM}}}{\dif t}=\frac{e^2 }{8\pi\lambda}\left|\frac{\mu_{ij}}{\mu_{B}}\right|^2Z^{2}F^{2}_{Z}(q^2)\left[\frac{1}{t}\left(2\lambda+4M^2m_{i}^2+2A\Delta+2M^2\Delta+\Delta^2 \right) +\left( 2A+\Delta\right)+\frac{2M^2\Delta^2}{t^2}  \right]
\label{crosssection:general} 
\end{equation}
where $\mu_{ij}$ is the neutrino magnetic moment connecting the mass eigenstates of $\nu_{i}$ and $\nu_{j}$, $\mu_B$ is the Bohr Magneton, and 
\begin{equation}
	A=s-M^2-m_{i}^{2}\;,\quad \Delta=m_{i}^2-m_{j}^2\;,\quad\lambda=A^2-4M^2m_{i}^2\,,
\end{equation}
with $ s $ and $ t $ being the Mandelstam variables, and in the laboratory frame are given by
\begin{equation}
	s=M^2+m_{i}^2+2ME_{\nu}\;\quad{\rm and}\quad\;t=-2MT\;.
\end{equation}
Note that the neutrino mass eigenstates with ($i,j=1,2,3,4$) include three kinds of active neutrinos and one sterile neutrino $\nu_4$. 

In this work, we consider the CE$\nu$NS process with solar neutrinos scattered to either the active or heavy sterile neutrino eigenstates. Therefore, the matrix of neutrino magnetic moments $ \mu_{ij} $ can be written as 
\begin{equation}
    (\mu_{ij})=\left(\begin{array}{ccc|c}
         \mu_{11}&\mu_{12} &\mu_{13} &\mu_{14}  \\
         \mu_{21}&\mu_{22} &\mu_{23} &\mu_{24}  \\
         \mu_{31}&\mu_{32} &\mu_{33} &\mu_{34}  \\
         \hline
         \mu_{41}&\mu_{42} &\mu_{43} &\mu_{44}  \\
    \end{array}\right)\,,
\label{matirce:mm}
\end{equation}
where the top-left part is the matrix of the active neutrino magnetic moment ($\nu_{a}$MM) and the left-bottom and right top parts are the
active-sterile transition magnetic moment ($\nu_{s}$MM). The right bottom part $\mu_{44}$ 
is the diagonal magnetic moment in the elastic scattering of the sterile neutrino $\nu_4$.
Note that we have neglected the flavor mixing between active and sterile neutrinos, 
thus there is no sterile neutrino component in the solar neutrino flux, and $\nu_4$ can only be generated by neutrino upscattering via the $\nu_{s}$MM. 

In the CE$\nu$NS process considered in this work, the electron neutrino produced near the solar core is a superposition of three active neutrino mass states and the magnetic moment at the detection is an effective magnetic moment with the neutrino mixing and flavor oscillation considered, which is given by~\cite{Giunti:2014ixa,Balantekin:2013sda}
\begin{equation}
    \mu_{{\rm S}}^2(L, E_{\nu})=\sum_{l}\left|\sum^{3}_{k=1}(U^M_{ek})^*e^{-i\Delta m^{2}_{kl}L/2E_{\nu}}\mu_{lk}\right|^2\,,
\end{equation}
where $U^M_{ek}$ is the mixing matrix element at the neutrino production in the solar core,
and $\Delta m^{2}_{kl}\equiv m^2_{k}-m^2_{l}$ is the mass squared difference of neutrino mass eigenstates.
Since the distance $L$ between the Sun and the Earth is much larger than the oscillation lengths, the interference terms $e^{-i\Delta m^2_{kl}L/2E_{\nu}}$ are washed out during the measurement. Then the effective magnetic moment at the detection can be written as
\begin{equation}
    \mu_{{\rm S}}^2(E_{\nu})=\sum^{3}_{k=1}\left|U^M_{ek}\right|^2\sum_{l}\left|\mu_{lk}\right|^2\equiv\sum_{\alpha}\left|\mu_{\nu_{\alpha}}\right|^2\left[\sum^{3}_{k=1}\left|U^M_{ek}\right|^2\left|U_{\alpha k}\right|^2\right]\,,
\end{equation}
where $\mu_{\nu_{\alpha}}$ is the effective magnetic moment of the flavor neutrino $\nu_{\alpha}$.
Since the electron neutrinos produced in the solar core turn into the fluxes with $ \nu_e $, $ \nu_\mu $ and $ \nu_\tau $ through neutrino oscillations when arrive at the Earth, in this work we consider the electron neutrino magnetic moment $ \mu_{\nu_e} $ and an effective magnetic moment $ \mu_{\mu\tau}^{\rm eff} $ for $ \nu_{\mu} $ and $ \nu_{\tau} $, which can be expressed as
\begin{equation}
\left( \mu_{\nu_{\mu\tau}}^{\rm eff}\right)^2=
\frac{\mu^2_{\nu_{\mu}}|U_{\mu k}|^2+\mu^2_{\nu_{\tau}}|U_{\tau k}|^2}{|U_{\mu k}|^2+|U_{\tau k}|^2}
\simeq 0.49 \mu_{\nu_{\mu}}^2+0.51 \mu_{\nu_{\tau}}^2\,.
\end{equation}
In the solar neutrino CE$\nu$NS detection, the neutrino energy $E_{\nu}$ and the nuclear recoil energy $T$ are much larger than the present upper limit of the active neutrino masses and the relevant variable $\Delta$ can be neglected. Therefore, considered the effective magnetic moment $\mu_{\nu_{\alpha}}$ with flavor $ \alpha=e,\mu,\tau$ described above and Eq.~(\ref{crosssection:general}), the popular cross section with the active $ \nu_{\alpha} $ magnetic moment at low energies can be written as~\cite{Cadeddu:2020lky,PhysRevD.39.3378,Giunti:2014ixa,Yue:2021vjg}
\begin{equation}
\frac{\dif \sigma_{\nu_{\alpha}{\rm MM}}}{\dif T}=\frac{\pi \alpha_{\rm EM}^2}{m_e^2}Z^{2}F^{2}_{Z}(q^2)\left|\frac{\mu_{\nu_{\alpha}}}{\mu_{B}}\right|^2\left(\frac{1}{T}-\frac{1}{E_{\nu}}\right)\,,
\label{crosssection:AMM}
\end{equation}
where $ m_e $ is the electron mass, $T$ is the nuclear recoil energy, $E_{\nu}$ is the neutrino energy and $ \alpha_{\rm EM} $ is fine structural constant. $Z$ is the number of protons and $F_{Z}(q^2)$ is the Helm form factor for protons and only protons contribute to this scattering process.

Next for the $\nu_{s}$MM scenario, an active neutrino $\nu_{l}$ up-scattering to a heavy sterile neutrino $ \nu_{4} $ can be described with the $\mu_{l4}$ part in the Eq.(~\ref{matirce:mm}). In this work, we assume that all active neutrinos universally couple to $\nu_4$ through the dipole process and we define a universal magnetic moment in this process with the variable $d$ given by
\begin{equation}
    d^2=\frac{\pi\alpha_{\rm EM}}{m_e^2}\left|\frac{\mu_{\nu_{l4}}}{\mu_{B}}\right|^2\,,
    \label{nsMM} 
\end{equation}
where $\mu_{\nu_{l4}}$ is the active sterile transition magnetic moment in Eq.(~\ref{matirce:mm}),
and the parameter $d$ is usually used in literatures as the coefficient of the effective Lagrangian~\cite{Schwetz:2020xra,Magill:2018jla}
\begin{equation}
\mathcal{L}=d\bar{\nu}_{l}\sigma^{\mu\nu}\nu_{4}F_{\mu\nu}+{\rm h.c.}\,,
\end{equation}
where $\nu_{4}$ is the sterile neutrino field and $F_{\mu\nu}$ is the electromagnetic field strength tensor. Note that both $\mu_{\nu_{l4}}$ and $d$ describe the strength of the $\nu_{s}$MM and they are connected through Eq.~(\ref{nsMM}).
Taking Eq.~(\ref{crosssection:general}) and neglecting the active neutrino masses the cross section induced by the $\nu_{s}$MM at low energies can be written as~\cite{Brdar:2020quo,Shoemaker:2018vii}
\begin{equation}
\frac{\dif \sigma_{\nu_{s}{\rm MM}}}{\dif T}=d^2 \alpha_{\rm EM} Z^{2}F^{2}_{Z}(q^2)\left(\frac{1}{T}-\frac{1}{E_{\nu}}-\frac{m_4^2}{2 E_{\nu} MT}\left(1-\frac{M-T}{2E_{\nu}}\right)-\frac{m_4^4(M-T)}{8E_{\nu}^{2}M^2T^2}\right)\,.
\label{crosssection:NDP} 
\end{equation}
Note that this cross section can return to the popular form of Eq.~(\ref{crosssection:AMM}) when $m_4=0$ is taken.

\section{Numerical Framework}

In this section, we are going to present the statistical method and setup of the simplified experimental scenarios~\cite{Li:2022jfl} we have employed to analyze the sensitivity of next generation and future dark matter DD experiments to the $\nu_{a}$MM and $\nu_{s}$MM.

\subsection{Statistical Method}
In this work we discuss the solar neutrino CE$ \nu $NS process in the presence of the $\nu_{a}$MM and $\nu_{s}$MM. The predicted CE$ \nu $NS event rate $ N_{i} $ in each nuclear-recoil energy-bin is given by
\begin{equation}
N_{i} = \frac{\epsilon}{M} \int_{T_{{i,\rm min}}}^{T_{{i,\rm max}}} \dif{T} \int_{E_{{\rm min}}}^{E_{{\rm max}}} \dif{E_{\nu}} \cdot  \Phi (E_{\nu}) \frac{\dif{\sigma}}{\dif{T}}\,,
\end{equation}
where $\epsilon$ is the exposure of the considered experiment and $M$ is the mass of the target nucleus, $\Phi(E_{\nu})$ is the solar neutrino fluxes from the standard solar model BS05(OP) (SSM)~\cite{Bahcall:2004mq,Bahcall_2005,Essig:2018tss}. $E_{{\rm max}}$ is the maximal neutrino energy from the solar neutrino spectrum. $E_{{\rm min}}$ the minimal neutrino energy for a certain recoil energy $T$. For the CE$ \nu $NS process with the $\nu_{a}$MM, it is given by
\begin{equation}
E_{{\rm min}}=\frac{T}{2}\left(1+\sqrt{1+2\frac{M}{T}}\right)\,.
\label{Emin_amm} 
\end{equation}
For the $\nu_{s}$MM case, $E_{{\rm min}}$ is higher due to the final heavy neutrino state $ \nu_4 $ with a mass $ m_4 $ and can be written as
\begin{equation}
E_{{\rm min}}=\frac{m_4^2+2MT}{2\left[\sqrt{T(T+2M)}-T\right]}
\label{Emin_ndp} \,.
\end{equation}

To explore the constraints on the $\nu_{a}$MM and $\nu_{s}$MM from the solar neutrino CE$\nu$NS process with certain DD experiments, we employ the standard least squares method with the Asimov data sets~\cite{Cowan:2010js}
\begin{equation}
\chi^2=\sum_{i=1}^{n}\left(\frac{N_{i}^{\rm exp}-(1+\epsilon_{\rm exp}) N_{i}^{\rm pred}[(1+\epsilon_{j})\Phi^{j}_{\rm SSM}]}{\sigma_{i}}\right)^{2}+\left(\frac{\epsilon_{\rm exp}}{\sigma_{\rm exp}}\right)^{2}+\Sigma_{j}\left(\frac{\epsilon_{j}}{\sigma_{\Phi^{j}_{\rm SSM}}}\right)^{2}\,,
\end{equation}
where $\sigma_{i}^2=N_{i}^{\rm exp}+N_{i}^{\rm bg}$, 
$N_{i}^{\rm exp}$ is the pseudo event number of the signal of the considered experiment in the $i$th energy bin, $N_{i}^{\rm pred}$ is the predicted event number, and $N_{i}^{\rm bg}$ is the number of background events. $\epsilon_{\rm exp}$ is the simplified nuisance parameter which quantifies the total detection uncertainty of the experiment and $\sigma_{\rm exp}$ is the corresponding standard deviation. $\epsilon_{j} $ and $ \sigma_{\Phi^{j}_{\rm SSM}} $ are the nuisance parameter and uncertainty of the $j$th solar neutrino flux from the SSM, in which the largest one is $11.6\%$ for the $^8{\rm B}$ neutrino flux. All the nuisance parameters will be varied to minimize the $\chi^2$ function. It is noteworthy that the quenching effect connecting the nuclear recoil to electron recoil response will be employed as described in the following subsection.

\subsection{Experimental Scenarios}

Today the DD experiments are entering the phase of the multi-ton scale and there are considerable experiments have the potential for the neutrino CE$\nu$NS detection and to explore the neutrino non-standard interactions. In the following, we summarize experimental scenarios with various target materials and detection technologies based on next generation and future DD experiments:

\begin{itemize}
\item Firstly, DD experiments with the liquid noble gas as the target have already achieved considerable results of the dark matter searches ~\cite{XENON100:2010cgk,LUX:2013afz,PandaX-II:2016vec,BKar,DEAP3600}. Xenon-based experiments including PandaX-4T~\cite{PandaX:2018wtu}, XENON-nT~\cite{XENON:2020kmp}, and LZ~\cite{LZ:2019sgr} have already deployed targets of the multiple ton scale and been able to observe the CE$\nu$NS process with solar neutrinos. In the future, DARWIN~\cite{DARWIN:2016hyl} will be able to deploy an unprecedented scale of target with 50 tons of liquid xenon and is promising to provide improved potential on WIMPs as well as the neutrino CE$\nu$NS process. Argon-based experiments are capable of detecting neutrinos with relatively lower energies due to the lighter mass of the argon nucleus and a higher production of liquid argon makes a larger scale of the detection target possible. Darkside-20k~\cite{DarkSide-20k:2017zyg} is planed to deploy 40 ton liquid argon for the dark matter and solar neutrino detection and ARGO~\cite{Billard:2021uyg} will increase the mass of liquid argon target to 400 tons in the future.

\item Secondly, low threshold dark matter detectors~\cite{CDMS:2013juh,SuperCDMS:2014cds,Aramaki:2016spe,CDEX:2018lau,DAMIC:2021esz,SENSEI:2020dpa} with the low scale solid material target can give constraints on low mass WIMPs. Next generations of this kind experiments, including Super CDMS~\cite{SuperCDMS:2016wui}, EDELWEISS-III~\cite{EDELWEISS:2017uga} and SENSEI~\cite{Tiffenberg:2017aac}, have the potential of detecting solar neutrino CE$\nu$NS events with extremely low recoil energies and may provide more severe tests for new physics beyond the SM.
\end{itemize}

\begin{table}
	\centering
	\begin{tabular}{ccccc}
		\toprule
		Type& Target & Exposure &   Threshold & Background  \\
		& & [t$\times$year] & [keV$_{\rm NR} $] &[t$^{-1}$year$^{-1}$keV$^{-1}$]\\
		\midrule 
		Ge-Next Generation &Germanium&0.2&0.1 &1 \\
		Si-Next Generation&Silicon&0.2 &0.1 &1 \\
		Xe-Next Generation&Xenon&20 &3.5&2 \\
		Ar-Next Generation&Argon&200 &3.5&2\\
		Ge-Future&Germanium&2&0.04&1 \\
		Si-Future&Silicon&2 &0.04 &1 \\
		Xe-Future&Xenon&200 &1&2 \\
		Ar-Future&Argon&3000 &1&2 \\
		\bottomrule
	\end{tabular}
	\caption{Experimental scenarios and their typical parameters employed in this work.}
	\label{table:exp}
\end{table}

Based on the above investigation, we have designed the experimental scenarios listed in Table~\ref{table:exp} with four target materials, where Next Generation indicates the experiments in the coming years and Future represents those in the far future with the comprehensively improved detection technologies.
In the table, the detection exposure, the energy threshold, and the background level are provided for each experimental scenario.
For each scenario, we consider a nominal systematic uncertainty of 10$ \% $ and an optimistic systematic uncertainty of 5$ \% $ to show the effects of different systematic uncertainties on the sensitivity results. We use a simplified flat background level for an economic computing resource budget based on the consideration in Refs.~\cite{SuperCDMS:2016wui,BKar,BKxe}, which are also listed in Table~\ref{table:exp}. The upper bound for our analysis is 10  keV$_{\rm NR}$ and signals submerge in background above this bound.

\begin{table}
	\centering
	\begin{tabular}{ccc}
		\toprule
		Target&Recoil energy  T[eV$_{\rm NR}$]  & Quenching factor $Y$\\
		\midrule 
		Liquid argon~\cite{COHERENT:2020iec}&1000 - 10000&$ 0.246+(7.8\times10^{-4}$ keV$_{\rm NR}^{-1} )T $\\
		\hline
		Liquid xenon~\cite{Wang:2016obw}&1000 - 10000&$ Y_{L}(T) $, $ k=0.133Z^{2/3}A^{-1/2} $\\
		\hline
		\multirow{2}{*}{Germanium~\cite{Essig:2018tss}}&40 - 250&$ 0.18\left[1-e^{-(T[{\rm{eV}}_{\rm NR}^{-1}]-15)/71.3} \right]  $\\
		\multirow{2}{*}{}&250 - 10000&$ Y_{L}(T) $, $ k=0.2 $\\
		\hline
		\multirow{2}{*}{Silicon~\cite{Essig:2018tss}}&40 - 254&$ 0.18\left[1-e^{-(T[{\rm{eV}}_{\rm NR}^{-1}]-15)/71.03} \right]  $\\
		\multirow{2}{*}{}&254 - 10000&$ Y_{L}(T) $, $ k=0.15 $\\
		\bottomrule
	\end{tabular}
	\caption{Quenching factors employed in this work.}
	\label{table:QF}
\end{table}

In the CE$\nu$NS process, the recoil energy of a nucleus will only be partly converted into the ionization energy $ E_{\rm ion} $~\cite{Essig:2018tss}, which can be directly detected in DD experiments and this quenching effect can be illustrated as
$E_{\rm ion}=Y T$, where $ T $ is the nuclear recoil energy and $ Y $ is the quenching factor. At high energies, the quenching factor can be theoretically estimated by the
Lindhard model~\cite{osti_4701226} as 
\begin{equation}
Y_{L}(T)=\frac{kg(\epsilon)}{1+kg(\epsilon)}\,,
\end{equation}
where 
$g(\epsilon)=3\epsilon^{0.15}+0.7\epsilon^{0.6}+\epsilon$ and $\epsilon=11.5Z^{-7/3}T$,
$ Z $ is the proton number and the nuclear recoil energy $ T $ is given in keV. The original theory by Lindhard sets $ k=0.133Z^{2/3}A^{-1/2} $,  where $ A $ is the mass number of the nucleus, but experimental data indicate a range of values under different recoil energies. For silicon and germanium targets, fitting results for the quenching factor from Ref.~\cite{Essig:2018tss} with high ionization efficiency are employed and $ k $ is set to 0.15 and 0.2 respectively for high recoil energies. In contrast, the linear quenching factors are used  for smaller recoil energy regions.
For the liquid argon target, we have adopted the linear quenching factor from simultaneous fits based on CENNS-10 from Ref.~\cite{COHERENT:2020iec}, and for the liquid xenon target we use the original Lindhard model~\cite{Wang:2016obw}. The expressions for quenching factors employed in this work are listed in Table.~\ref{table:QF}.

\section{Numerical Results}
In this section we present numerical analysis results. First, we show the predicted event spectra for each experimental scenario as a function of the equivalent electron recoil energy. Then we illustrate the sensitivity of the solar neutrino CE$ \nu $NS detection on the $\nu_{a}$MM and $\nu_{s}$MM in different experimental scenarios. We have also considered the general case with both the $\nu_{a}$MM and $\nu_{s}$MM to show what are the remaining parameter space with the solar neutrino CE$ \nu $NS detection.


\subsection{Predicted Event Spectra}

\begin{figure}
	\centering
	\includegraphics[scale=0.38]{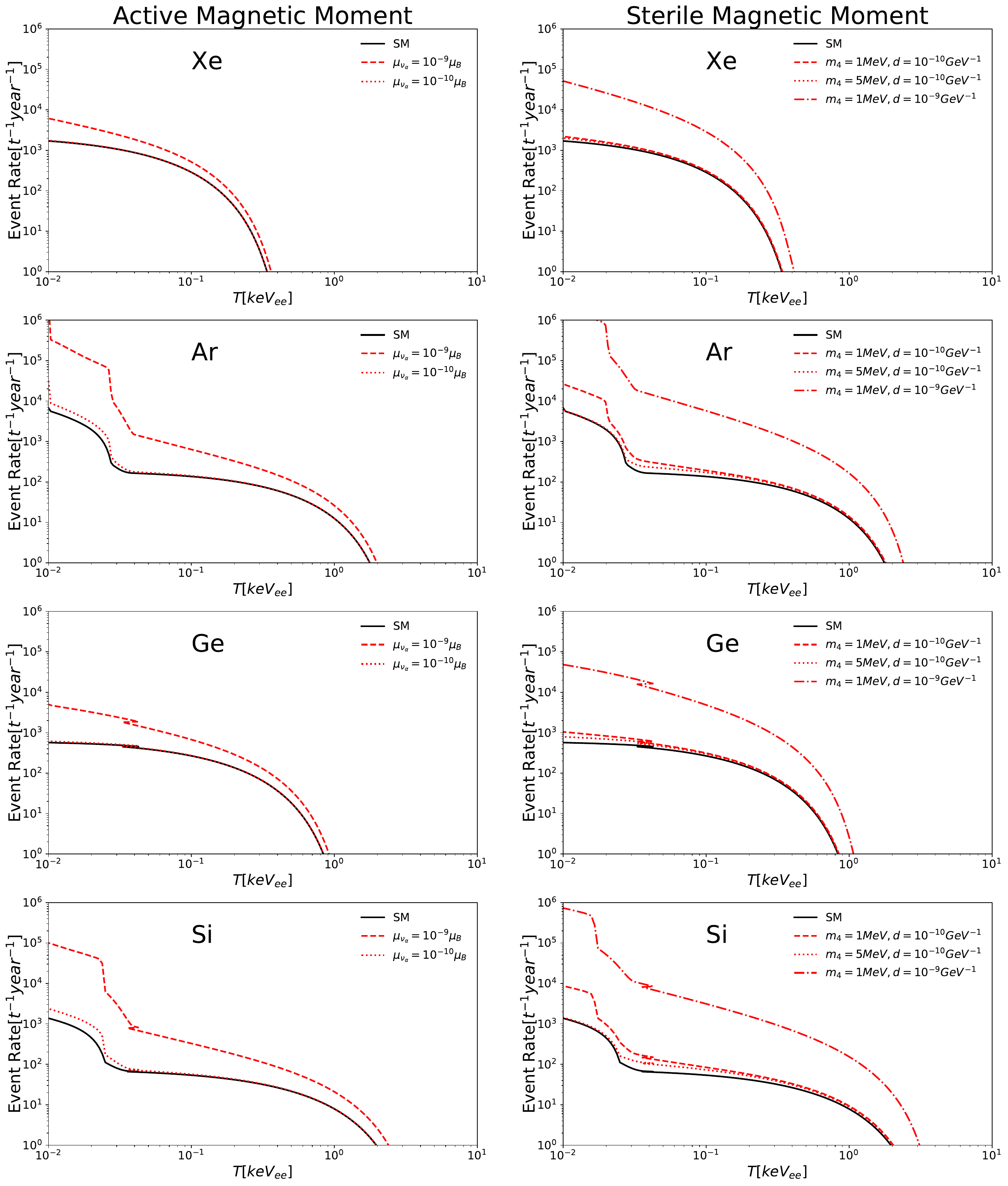}
	\caption{Predicted event energy spectra as a function of the nuclear recoil ionization energy for different detector materials in the presence of the $\nu_{a}$MM (left panel) and $\nu_{s}$MM (right panel) and with the quenching effect. A weighted average have been performed according to the natural abundance of isotopes in the detector material. From top to bottom rows results are shown for xenon, argon, germanium and silicon detectors respectively.}
	\label{fig:spectra}	
\end{figure} 

In Figure~\ref*{fig:spectra} we show the predicted event energy spectra as a function of the nuclear recoil ionization
energy for different detector materials in the presence of $\nu_{a}$MM (left panel) and $\nu_{s}$MM (right panel) and with the quenching effect. A weighted average have been performed according to the natural abundance of isotopes in the detector material. From top to bottom rows results are shown for xenon, argon, germanium and silicon detectors respectively. 
Since both contributions of the $\nu_{a}$MM and $\nu_{s}$MM are suppressed by the recoil energy $ T $ as shown in Eq.~(\ref{crosssection:AMM}) and Eq.~(\ref{crosssection:NDP}), these effects would become significant when the recoil energy decreases and low threshold experiments are needed for improved exploration. Several comments on the spectra are provided as follows.
\begin{itemize}
\item For the $\nu_{a}$MM case, we have shown the relevant contribution with the neutrino magnetic moment $ \mu_{\nu_{l}} $ of all flavors set to $ 10^{-9}\mu_{B}$ and $ 10^{-10}\mu_{B} $ respectively.
The contribution of the $\nu_{a}$MM is proportional to the square of $ \mu_{\nu_{l}} $ and 
the magnetic moment of $ 10^{-10}\mu_{B} $ or smaller is extremely hard to probe even with the 10 eV threshold for xenon and germanium targets. Lighter nuclei like argon and silicon have visible sensitivity for the 10 eV threshold due to relatively high recoil energies.

\item The contribution of the $\nu_{s}$MM is similar to that of the $\nu_{a}$MM case while the mass of the sterile neutrino significantly suppresses the event rates due to the minimal energy required for solar neutrinos to up-scatter a heavy sterile one as shown in Eq.(~\ref{Emin_ndp}). Therefore, effects of the $\nu_{s}$MM are completely suppressed at low recoil energies as shown in several tens of eV in the energy spectra with argon and silicon as targets. For liquid xenon and germanium targets this phenomenon will happen at the lower recoil energies because of heavier nuclear masses.
\end{itemize}

\subsection{Constraints on the $\nu_{a}$MM and $\nu_{s}$MM}

\begin{figure}
	\centering
	\includegraphics[scale=0.23]{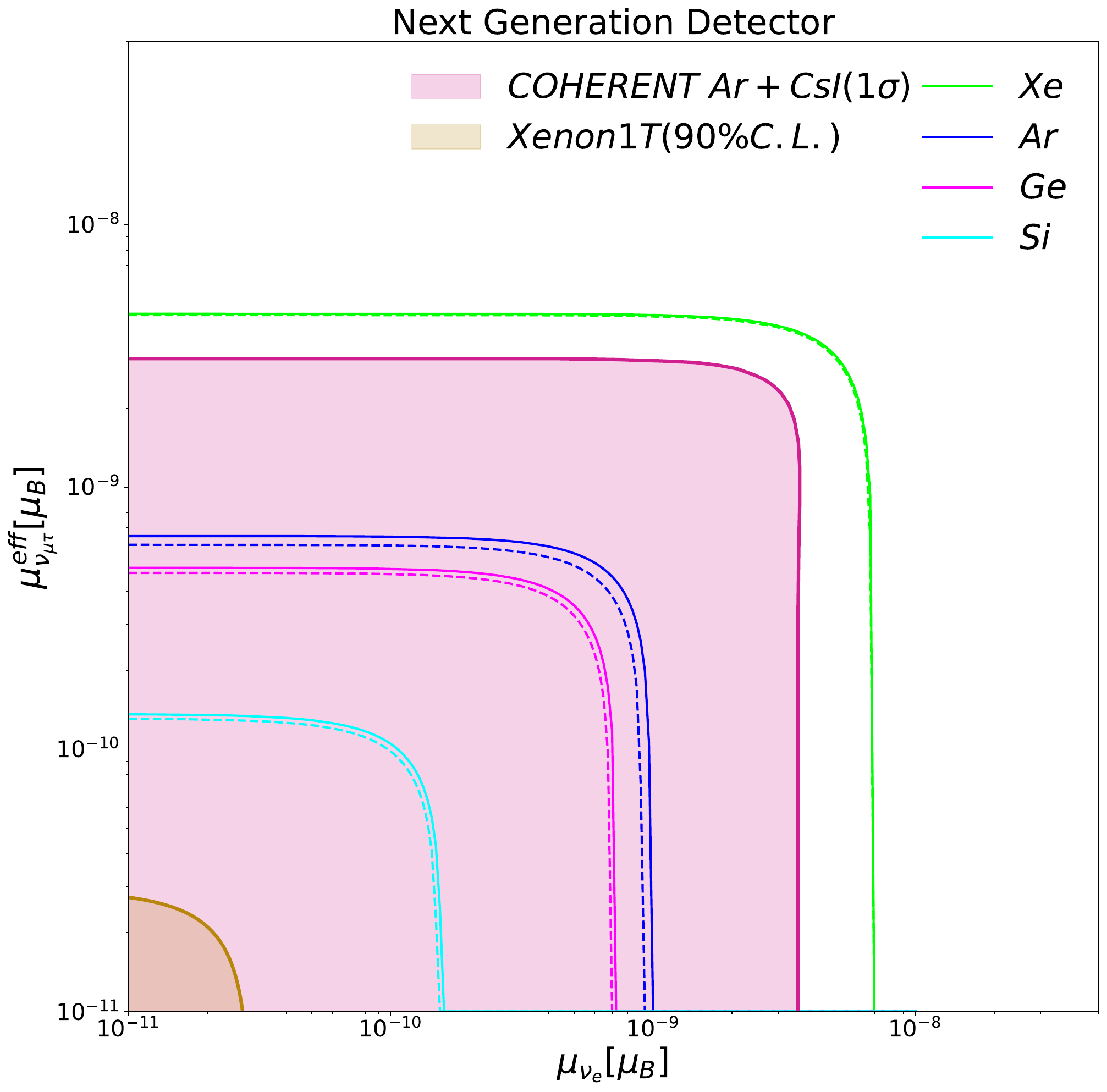}
	\includegraphics[scale=0.23]{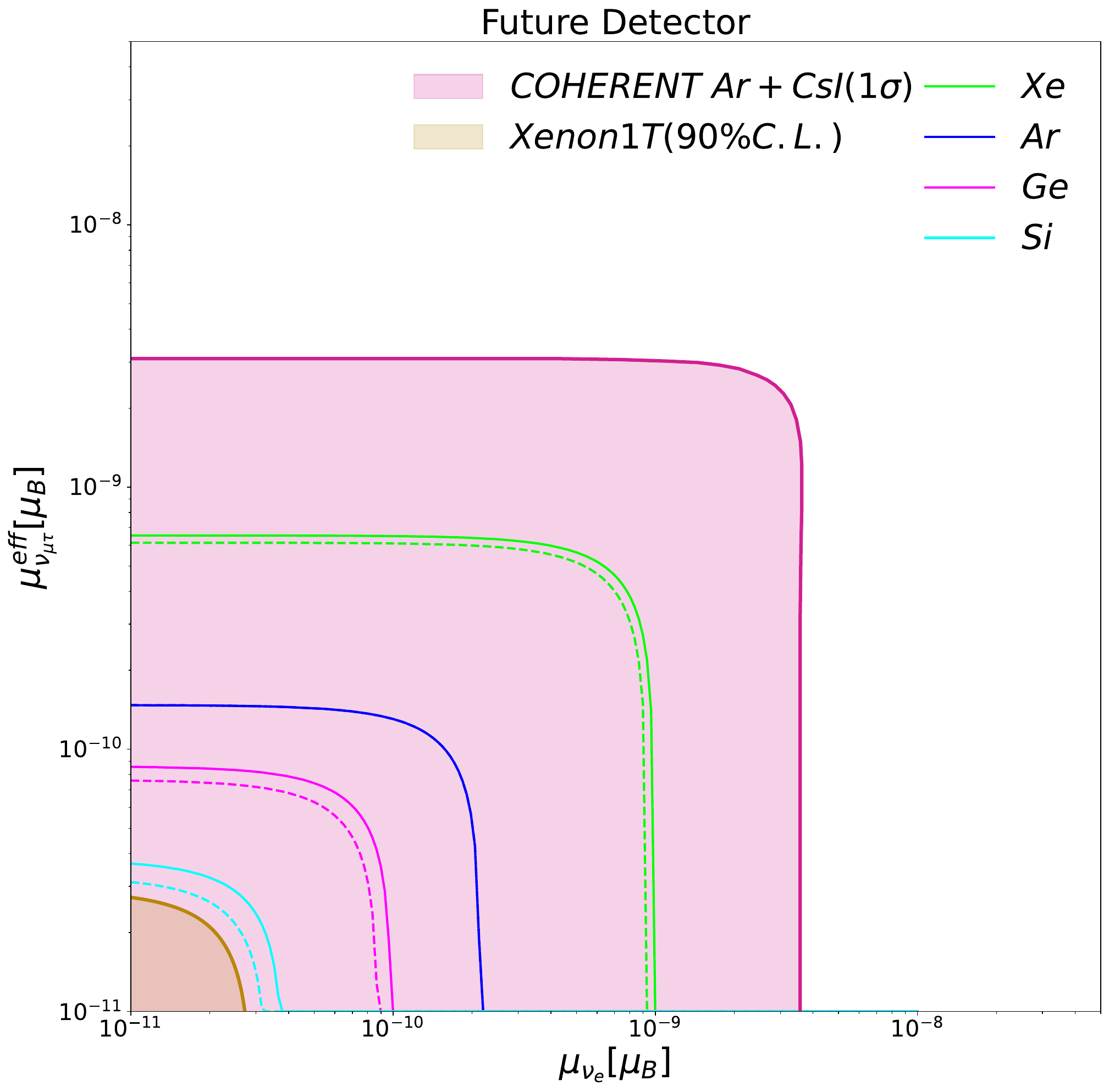}
	\caption{90$\%$ confidence level (C.L.) upper limits on the parameter space of $\nu_{a}$MM from the experimental scenarios listed in Table.~\ref{table:exp},where the solid lines are for the nominal systematic uncertainty and dashed lines for the optimal systematic uncertainty. The left and right panels are shown for the next generation and future experimental scenarios, respectively. A weighted average have been performed according to the natural abundance of isotopes in detector material. The allowed region from the Xenon1T excess~\cite{XENON:2020rca} and constraint from the combined analysis of COHERENT Ar and CsI data~\cite{Cadeddu:2020lky} are also shown for comparison.}
	\label{constraints:AMM}
\end{figure}

\begin{figure}
	\centering
	\includegraphics[scale=0.21]{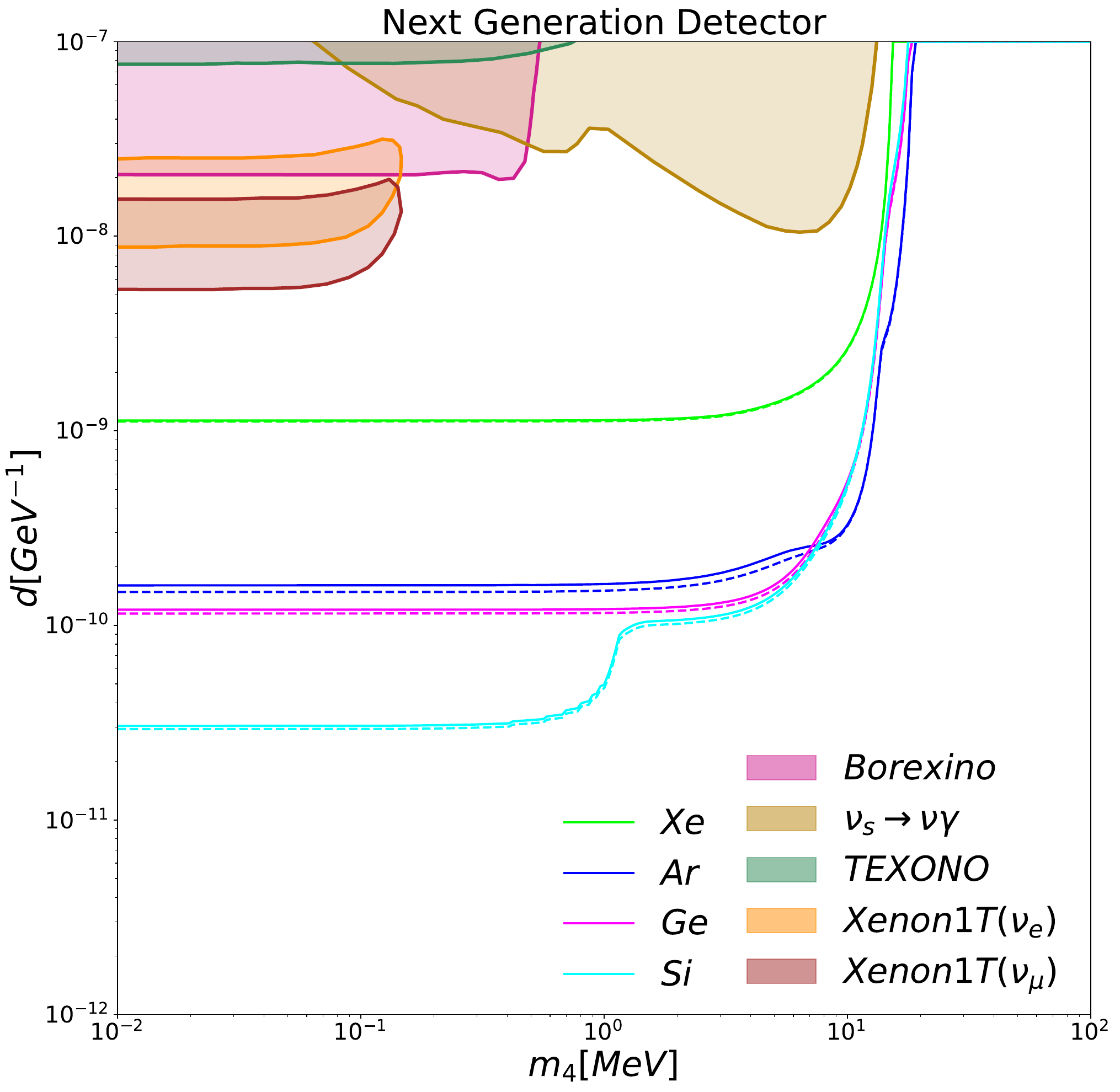}
	\includegraphics[scale=0.21]{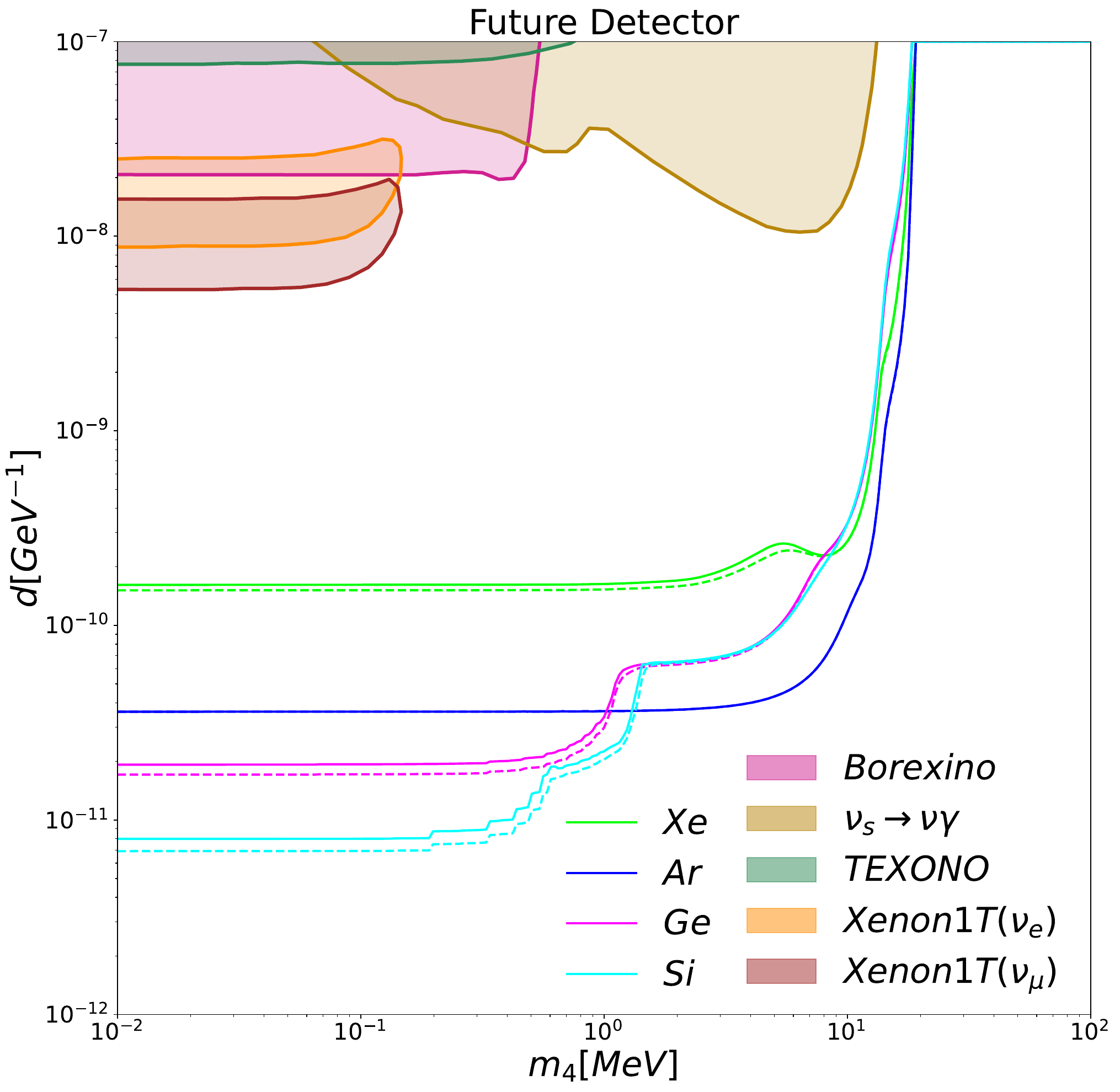}
	\caption{90$\%$ confidence level (C.L.) upper limits on the $\nu_{s}$MM parameter space from the experimental scenarios listed in Table.~\ref{table:exp},where the solid lines are for the nominal systematic uncertainty and dashed lines for the optimal systematic uncertainty. The left and right panels are shown for the next generation and future experimental scenarios, respectively. A weighted average have been performed according to the natural abundance of isotopes in detector material. 
	The allowed region from the Xenon1T excess~\cite{XENON:2020rca}, and the present experimental constraints from Borexino~\cite{Brdar:2020quo}, TEXONO~\cite{Miranda:2021kre,TEXONO:2009knm} and the $ \nu_{s}\rightarrow\nu\gamma $ decay sensitivity from Borexino and Super Kamiokande~\cite{Plestid:2020vqf} are shown for comparison.}
	\label{constraints:NDP}
\end{figure}

In Figure ~\ref{constraints:AMM} and Figure ~\ref{constraints:NDP} we have illustrated the 90\% confidence level (C.L.) upper limits on the parameter space of the $\nu_{a}$MM and $\nu_{s}$MM with the experimental scenarios listed in Table ~\ref{table:exp} respectively. The solid lines are for the nominal systematic uncertainty and dashed lines for the optimal systematic uncertainty. The left and right panels are shown for next generation and future experimental scenarios respectively and the natural abundance of isotopes has been taken into consideration with corresponding weighted average. The allowed regions of the Xenon1T excess at 90\% C.L. are illustrated as colored bands. Meanwhile the constraints from the combined analysis of COHERENT Ar and CsI data~\cite{Cadeddu:2020lky}, from Borexino~\cite{Brdar:2020quo}, $\nu_{s}\to\nu\gamma$~\cite{Plestid:2020vqf}, TEXONO~\cite{Miranda:2021kre,TEXONO:2009knm}
are also shown for comparison.

It is shown in Figure ~\ref{constraints:AMM} and Figure ~\ref{constraints:NDP} that the silicon target provides significantly more stringent constraints than others for the lightest nuclear mass and lowest detection threshold. Constraints from experiments with liquid argon and germanium as targets are more stringent than that of xenon-based experiments for higher exposure and better threshold respectively. It can be revealed that the contribution of better systematic uncertainty only take effects with enough exposure and threshold, which is the case of argon-based next generation experiments and future experiments with xenon, germanium and silicon as targets. However, improving systematic uncertainty may suffer from marginal effects when the threshold and statistics reach a certain level as shown in argon-based future experiments.
Some relevant comments are summarized as follows:
\begin{figure}
	\centering
	\includegraphics[scale=0.5]{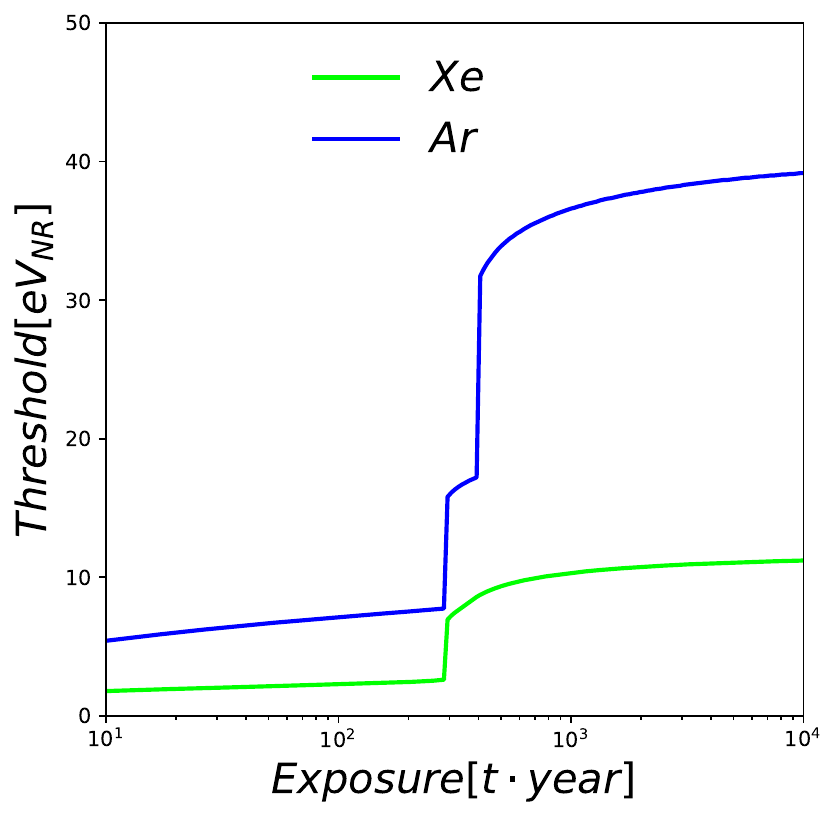}
	\includegraphics[scale=0.5]{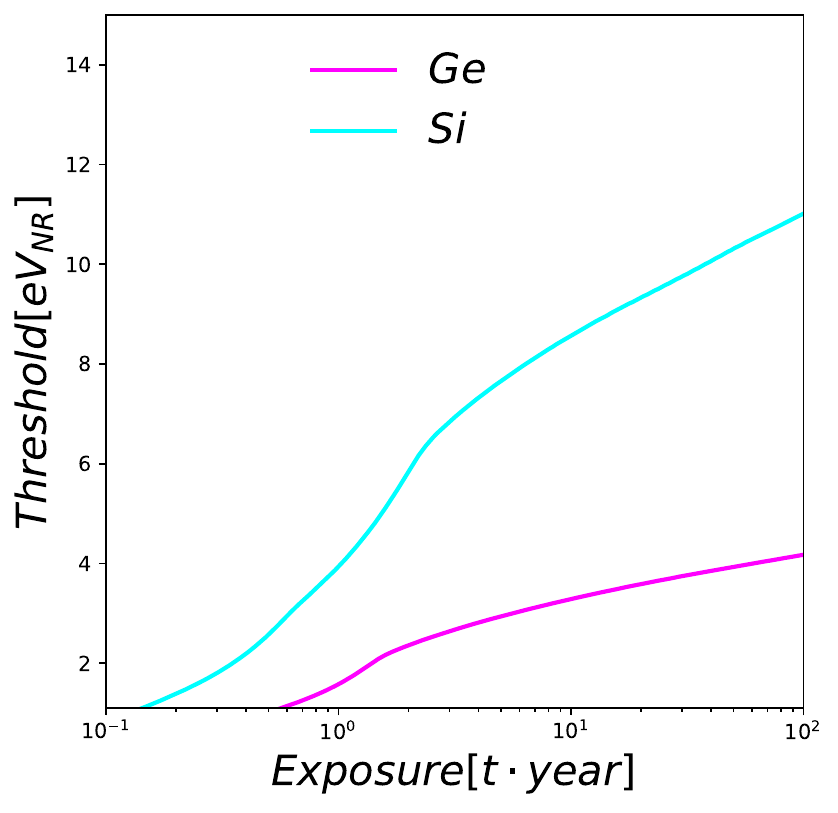}
	\caption{Requirements of different target materials to reach the $\nu_{a}$MM sensitivity with $\mu_{\nu_{l}}=10^{-11}\mu_B$ (90\%C.L.) as the function of the exposure and detection threshold.}
	\label{req:detector}
\end{figure}
\begin{itemize}
\item For the $\nu_{a}$MM, the next generation experiments give constraints at the level of $ \mathcal{O}(10^{-10})\mu_{B} $ except for xenon-based experiments, which are better than the limit from the combined analysis of COHERENT Ar and CsI data~\cite{Cadeddu:2020lky}. Future experimental scenarios provide significantly more stringent constraints and liquid noble gas based and low threshold solid based experiments are capable of the $\mu_{\nu_{l}}$ limit to the $ \mathcal{O}(10^{-10})\mu_{B} $ and $ \mathcal{O}(10^{-11})\mu_{B} $ levels respectively. It should be noticed that only future experiments with silicon as the target have the potential to exclude the parameter space to explain the excess from electron recoil results of Xenon1T~\cite{XENON:2020rca}.

\item It is shown in Figure ~\ref{constraints:NDP} that the CE$ \nu $NS detection in DD experiments provides excellent constraints on the $\nu_{s}$MM model, and both next generation and future experimental scenarios can unambiguously test the parameter space from the Xenon1T excess~\cite{XENON:2020rca}. Meanwhile, constraints from these experiments are significantly better than present experimental constraints from Borexino~\cite{Brdar:2020quo}, TEXONO~\cite{Miranda:2021kre,TEXONO:2009knm} and the $ \nu_{s}\rightarrow\nu\gamma $ decay sensitivity from Borexino and Super Kamiokande~\cite{Plestid:2020vqf}. Note that there are several sudden changes in the sensitivity curves of silicon-based experiments and future germanium-based experiments, which are resulted from the appearance of different low-energy solar neutrino fluxes. With the recoil energy threshold, the target nuclear mass and sterile neutrino mass considered, it can be easily concluded from the minimal neutrino energy to trigger the $\nu_{s}$MM process in Eq.~(\ref{Emin_ndp}) that the lower sudden changes of future silicon-based curves are from the appearance of the $^7$Be neutrino flux and the presence of CNO neutrino fluxes leads to the other sudden changes. Meanwhile, the $^8$B neutrino flux contributes to the constraints with higher sterile neutrino masses, but the $ pp $ and 0.384 MeV $ ^7 $Be neutrino fluxes are not able to provide the constraints beyond 10 eV due to lower recoil energies.

\item Comparing between Figure ~\ref{constraints:AMM} and Figure~\ref{constraints:NDP}, one can conclude that the CE$ \nu $NS process in DD experiments provides far more stringent constraints on the $\nu_{s}$MM than the electron neutrino elastic scattering (E$ \nu $ES) while weaker for the case of the $\nu_{a}$MM. 
In the presence of the $\nu_{a}$MM, E$ \nu $ES induces significantly more events than CE$ \nu $NS with advantages of both higher event rate at most recoil energies and a much larger allowed recoil energy range. In the presence of the $\nu_{s}$MM, the situation becomes more complicated since the minimal neutrino energies to trigger the up-scattering are not monotone increasing. It can be easily observed from Eq.~(\ref{Emin_ndp}) that the minimal neutrino energy $E_{{\rm min}}$ continuously decreases as the increase of the recoil energy $T$,
until $E_{{\rm min}}$ reaches the extreme value at $ T_{}=m_4^2/2(m_4+M)$, where $ M $ is the mass for the target electrons or nuclei and $m_4$ is the sterile neutrino mass. 
After the minimal neutrino energy reaches the extreme value, it will start to increase as the recoil energy. 
For the CE$ \nu $NS, the recoil energy $T$ at the minimal $E_{{\rm min}}$ is much smaller than the expected experimental detection threshold due to the heavy nuclear mass, and the minimal neutrino energy to trigger scattering is monotone increasing with the recoil energy in the measured energy range. For the E$ \nu $ES, since $ m_4 $ can be similar to the order of the electron mass, $T_{}$ at the minimal $E_{{\rm min}}$ will be in the detection energy range and may even exceeds the upper bound of the measured energy range when $ m_4 $ is at the level of several hundreds of keV. This kinetic constraints in the upscattering process with the $\nu_4$ and electron will limit the sensitivity of the E$ \nu $ES process to the $\nu_{s}$MM.
Therefore, it can be concluded that CE$ \nu $NS has significant advantages in constraining the $\nu_{s}$MM compared to the E$ \nu $ES process, especially for heavier sterile neutrinos.

\item It has been shown in Figure~\ref{constraints:AMM} that the constraints of the $\nu_{a}$MM can hardly reach the parameter space of the Xenon1T excess using the considered experimental scenarios. Therefore in Figure~\ref{req:detector} we further illustrate the requirements of different target materials to reach the $\nu_{a}$MM sensitivity with $\mu_{\nu_{l}}=10^{-11}\mu_B$ (90\%C.L.) as functions of the exposure and detection threshold. It is clearly that within the feasible exposure range at present or in the near future, extremely low threshold is needed to efficiently constrain the $\nu_{a}$MM compared to the E$ \nu $ES experiments. However, the atomic effects studied in Refs.~\cite{Ge:2021snv,Roberts:2016xfw,Catena:2019gfa,Kouzakov:2011vx}
may significantly affect the results, which need further research in the future. An extremely large exposure may also help to keep the threshold in a practical range, which is considered in Ref.~\cite{Ye:2021zso}.

\end{itemize}

\begin{figure}
	\centering
	\includegraphics[scale=0.23]{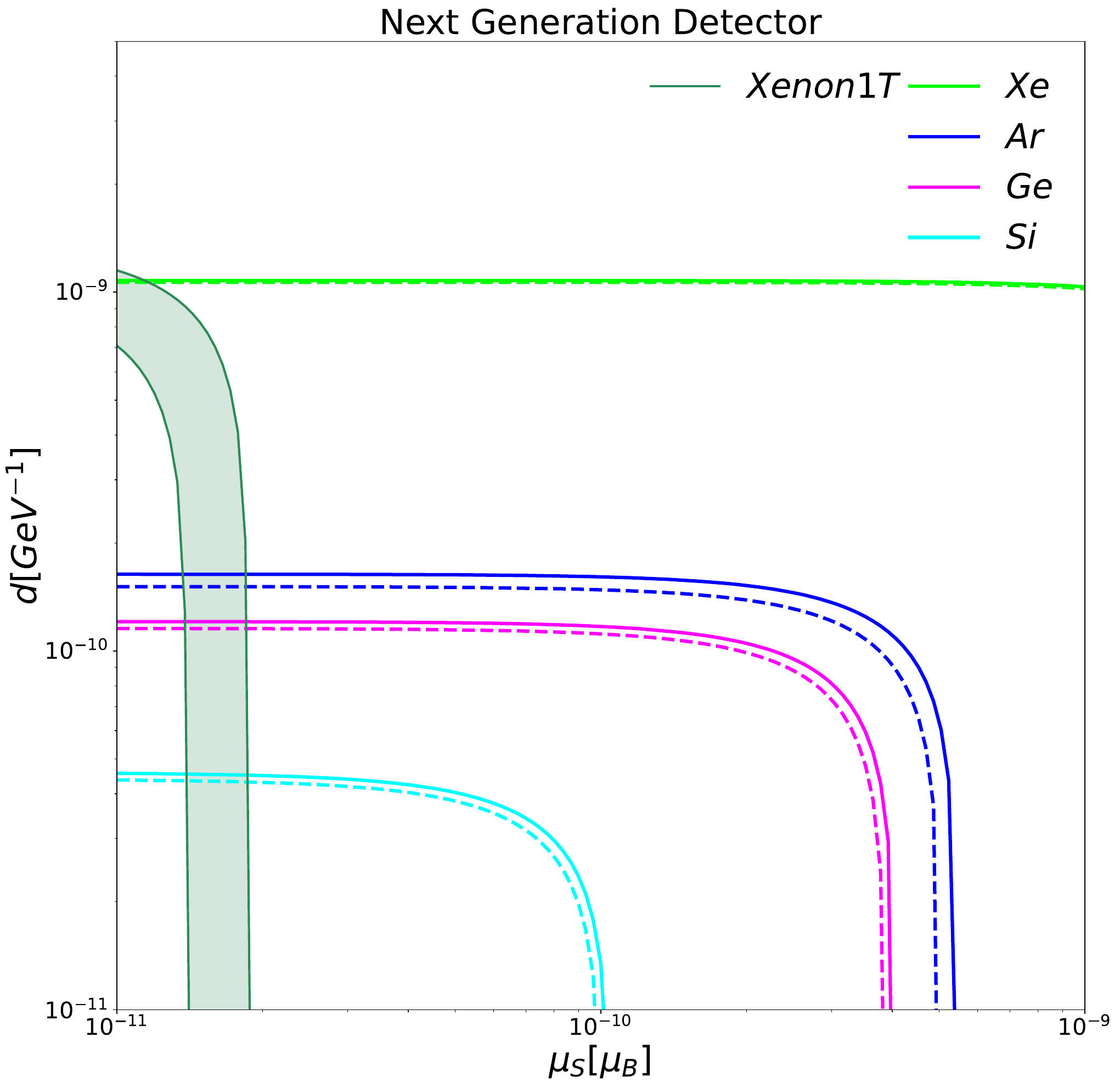}
	\includegraphics[scale=0.23]{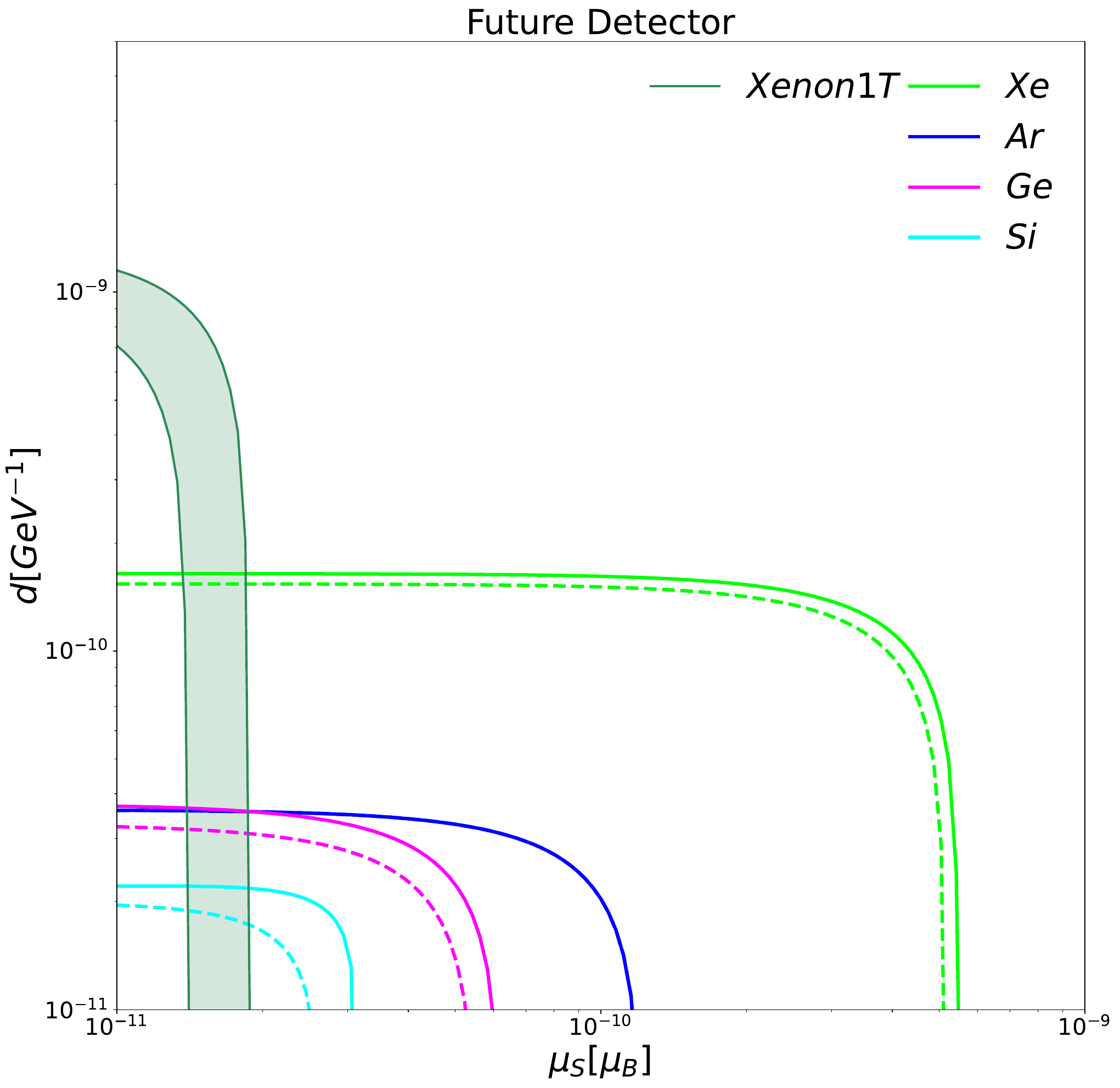}
	\caption{90$\%$ C.L. upper limits on the neutrino magnetic moments in the presence of both the $\nu_{a}$MM, denoted as $\mu_{\rm S}$, and the $\nu_{s}$MM in term of $d$ with $ m_4=1$ MeV from the experimental scenarios listed in Table.~\ref{table:exp}, where the solid lines are for the nominal systematic uncertainty and dashed lines for the optimal systematic uncertainty. The left and right panels are shown for the next generation and future experimental scenarios, respectively. A weighted average have been performed according to the natural abundance of isotopes in the detector materials. We also show the allowed parameter space to explain the Xenon1T excess~\cite{XENON:2020rca}.}
	\label{constraints:mix}
\end{figure}
Finally, we study the general framework in the presence of both the $\nu_{s}$MM and $\nu_{a}$MM and illustrate the sensitivity in Figure ~\ref{constraints:mix} by fixing $ m_4=1$ MeV. We show the constraints on both $\nu_{s}$MM and $\nu_{a}$MM with different experimental scenarios, together with the allowed parameter space to explain the Xenon1T
excess~\cite{XENON:2020rca}. It is observed that both next generation and future DD experiments except xenon-based ones can exclude the allowed range of the $\nu_{s}$MM with $ d \gtrsim 10^{-10}\, {\rm{GeV}}^{-1} $.
However, none of the experimental scenarios could exclude the region of $ d < 10^{-11}\, {\rm{GeV}}^{-1} $ where the $\nu_{a}$MM dominates the contribution to the Xenon1T
excess.
In this respect, other probes, such as the E$ \nu $ES process, are needed to test the explanation of the Xenon1T excess. 

\section{Conclusion}

The dark matter DD experiments are entering the multi-ton scale phase, which also have great potential to detect the solar neutrino CE$\nu$NS process. In this work, we have presented the detection potential of the $\nu_{a}$MM and $\nu_{s}$MM in the solar neutrino CE$\nu$NS process in the next generation and future DD experiments.
We have illustrated the sensitivity of the $\nu_{a}$MM and $\nu_{s}$MM, and compared with the respective allowed range of the Xenon1T excess and other experimental constraints.
We find that the $\nu_{a}$MM sensitivity lies in the levels of [$10^{-10}$, $10^{-11}$]$\mu_B$, which is dominantly limited by the detection threshold. Only future silicon based experiments could test the allowed region of the Xenon1T excess.
On the other hand, the solar neutrino CE$\nu$NS process is powerful to probe the $\nu_{s}$MM for the sterile neutrino mass below 10 MeV. 
All the considered scenarios can unambiguously test the $\nu_{s}$MM explanation of the Xenon1T excess with at least one order of magnitude better detection sensitivities.
Finally we have also derived the detection potential in the general framework with both $\nu_{a}$MM and $\nu_{s}$MM contributions in the CE$\nu$NS. Therefore we conclude that the future the solar neutrino CE$\nu$NS detection would be a new and promising tool to test new physics beyond the SM.


\section*{Acknowledgements}
YFL is grateful to Dr.~Jing-yu Zhu for helpful discussions on models of the neutrino dipole portal. The work of YFL and SYX was supported by National Natural Science Foundation of China under Grant Nos.~12075255, 12075254 and 11835013, by the Key Research Program of the Chinese Academy of Sciences under Grant No.~XDPB15.

\bibliographystyle{h-physrev5}
\bibliography{main}

\end{document}